\documentclass[fleqn,twoside]{article}
\usepackage{espcrc2}

\usepackage{graphicx}
\usepackage[figuresright]{rotating}

\newcommand{\AmS}{{\protect\the\textfont2
  A\kern-.1667em\lower.5ex\hbox{M}\kern-.125emS}}

\newcommand{\mbf}[1]{\mbox{\boldmath $#1$}}

\title{Diquark interaction and gaps for color superconductivity}

\author{E. Gubankova 
\address{Institute of Theoretical and Experimental Physics,\\
B. Cheremushkinskaya 25, RU-117 218 Moscow, Russia}\thanks{
Current address: Center for Theoretical Physics,
Massachusetts Institute of Technology, 
Cambridge, MA 02139, USA} }

\begin{document}

\begin{abstract}
Using flow equations, we derive an effective quark-quark 
interaction and obtain the coupled set of gap equations
for the condensates of the CFL phase of massless $N_f=3$
dense QCD. The formalism developed here enables one to consider
more general case of nonzero $s$-quark mass.
\vspace{1pc}
\end{abstract}

\maketitle

We apply flow equations to the Coulomb gauge QCD Hamiltonian with
$N_f=3$ at nonzero quark density. 
Coulomb gauge QCD Hamiltonian, $\nabla\cdot A =0$, at nonzero quark density
\begin{eqnarray}
H &=& H_{0}+H_{inst}+H_{dyn}
\,,\label{eq:1.1}\end{eqnarray}
includes the free Hamiltonian $H_{0}$, the instantaneous interaction $H_{inst}$
describing static properties, and the dynamical interaction $H_{dyn}$
involving gluon propagation. The free Hamiltonian is given by
\begin{eqnarray}
 H_{0} &=& \int d \mbf{x}\bar{\psi}(\mbf{x}) 
\left( -i \mbf{\gamma}\cdot\mbf{\nabla} -\mu\gamma_0
+ m \right)\psi(\mbf{x})
\nonumber\\ 
&+& {\rm Tr}\int d \mbf{x} \left( \mbf{\Pi}^2(\mbf{x})
+ \mbf{B}^2_{A}(\mbf{x}) \right)
\,.\label{eq:1.2}\end{eqnarray}
The physical degrees of freedom are the transverse gluon field $\mbf{A}$,
its conjugate momentum $\Pi$, and the quark field in the Coulomb gauge;
$\mbf{B}_{A}$ is the abelian component of the magnetic field.
The instantaneous interaction is given by
\begin{eqnarray}
 H_{inst} &=& -\frac{1}{2}\int d\mbf{x}d\mbf{y}
\rho^{a}(\mbf{x})V_{inst}^{ab}(\mbf{x},\mbf{y})
\rho^{b}(\mbf{y})
\,,\label{eq:1.3}\end{eqnarray} 
where the quark charge density is
$\rho^{a}(\mbf{x})=\bar{\psi}(\mbf{x})\gamma_0 T^a\psi(\mbf{x})$,
and in the leading order the kernel is diagonal 
$\sim\delta^{ab}$ Coulomb potential
$V_{inst}(r) = - \alpha_s/r$ in coordinate space or
$V_{inst}(q) = - g^2/\mbf{q}^2$ in momentum space.
In high density quark matter the non-abelian contributions from three-
and four-gluon interactions are suppressed. Therefore
we consider only the quark-gluon interaction, $H_{dyn}=V_{qg}$, 
\begin{eqnarray}
 V_{qg} &=& -g \int d \mbf{x}\bar{\psi}(\mbf{x})
\mbf{\gamma}\cdot\mbf{A}(\mbf{x}) \psi(\mbf{x})
\,,\label{eq:1.4}\end{eqnarray}
in dense QCD.

We eliminate the quark-gluon coupling  by the first order flow equation
$dV_{qg}/dl=[[H_{0},V_{qg}],H_{0}]$. As a result, in the second order
we generate the dynamical interaction between quarks 
$dV_{dyn}/dl=[[H_{0},V_{qg}],V_{qg}]_{two-particles}$,
and the quark self-energy 
$d\Sigma_{dyn}/dl=[[H_{0},V_{qg}],V_{qg}]_{one-particle}$
in the channel of $qq$-pair creation/annihilation.
The resulting diquark interaction and self-energy
include dynamical terms generated by flow equations,
$V_{dyn}(\Lambda\rightarrow 0)$ and $\Sigma(\Lambda)$, respectively,
and the instantaneous terms in corresponding channels, i.e.
$V_{qq}=V_{dyn}+V_{inst}$ and
$\Sigma_{qq}=\Sigma_{dyn}(\Lambda)+\Sigma_{inst}(\Lambda)+CT(\Lambda)$.
Here $\Lambda$ is the UV cutoff connected to the flow parameter $l$ as 
$l=1/\Lambda^2$, and $CT(\Lambda)$ are the second order counterterms.
Having the effective interaction between quarks $V_{qq}$, we allow
for diquark condensation, which can be parametrized as \cite{AlfordRajagopalWilczek}
$\Delta^{ij}_{\alpha\gamma}(\mbf{p})=3\left(\frac{1}{3}\left[\Delta_{8}(\mbf{p})
+\frac{1}{8}\Delta_{1}(\mbf{p})\right]\delta^{i}_{\alpha}\delta^{j}_{\gamma}
+\frac{1}{8}\Delta_{1}(\mbf{p})\delta^{i}_{\gamma}\delta^{j}_{\alpha}
\right)$, where $\Delta_1$ and $\Delta_8$ are the eigenvalues of
$\Delta^{ij}_{\alpha\gamma}$ in the CFL basis. Here
$i,j$ are flavor and $\alpha,\gamma$ color indices, which are substituted
by $\rho$ index in the CFL basis, 
$b(\mbf{k})^{i}_{\alpha}=\sum_{\rho}\lambda^{\rho}_{i\alpha}b(\mbf{k})^{\rho}/\sqrt{2}$,
$\lambda^{\rho}$ are the Gell-Mann matrices for $\rho=1,...,8$ and 
$\lambda^{9}=\sqrt{2/3}$. Combining all the terms
the high density effective Hamiltonian is given by \cite{Gubankova}
\begin{eqnarray}
H_{eff} &=& H_0+\Sigma_{qq}+V_{qq}
\,.\label{eq:1.5}\end{eqnarray}
Using Weyl spinors, the individual terms are written 
in the CFL basis as  
\begin{eqnarray}
 H_0 &=&  \sum_{\mbf{k},\rho}
|k-\mu|~b^{\dagger}_{\rho}(\mbf{k})b_{\rho}(\mbf{k})
+\sum_{\mbf{k}}k~a^{\dagger}(\mbf{k})a(\mbf{k})
\nonumber\\
 \Sigma_{qq} &=& \frac{1}{2}\sum_{\mbf{p},\rho}
\Delta_{\rho}(\mbf{p})
{\rm e}^{-i\phi(\mbf{p})}
b_{\rho}(\mbf{p})b_{\rho}(-\mbf{p})+c.c.
\nonumber\\
 V_{qq} &=& \sum_{\mbf{k},\mbf{p},\rho,\rho'}
V^{\rho\rho'}_{qq}(k,p)c^{\rho\rho'}
{\rm e}^{i\phi(\mbf{k})}{\rm e}^{-i\phi(\mbf{p})}
\nonumber\\
&\times&b^{\dagger}_{\rho}(\mbf{k})b^{\dagger}_{\rho}(-\mbf{k})
b_{\rho'}(\mbf{p})b_{\rho'}(-\mbf{p}) 
\,,\label{eq:1.6}\end{eqnarray}
where $\sum_{\mbf{k}}=\int d\mbf{k}/(2\pi)^3$, and
$c^{\rho\rho'}=1$ for $(\rho,\rho')=(1,8)$ or $(8,1)$,
$c^{\rho\rho'}=-2$ for $(\rho,\rho')=(8,8)$ and
$c^{\rho\rho'}=0$ for $(\rho,\rho')=(1,1)$. We did not
include antiparticles since their contribution is suppressed
near the Fermi surface. Performing the Bogoliubov-Valatin transformation
for quark fields from $(b,b^{\dagger})$ to $(y,y^{\dagger})$,
we absorb the self-energy into the new free Hamiltonian
\cite{AlfordRajagopalWilczek},
$H_0+\Sigma_{qq}\rightarrow \tilde{H}_0$,
\begin{eqnarray}
\tilde{H}_{0} &=& \sum_{\mbf{k},\rho}E_{\rho}(\mbf{k})
~y^{\dagger}_{\rho}(\mbf{k})y_{\rho}(\mbf{k})
\nonumber\\
&+& \sum_{\mbf{k}}\omega(\mbf{k})~a^{\dagger}(\mbf{k})a(\mbf{k})
\,,\label{eq:1.7}\end{eqnarray}
where the effective quark energy is given by
$E_{\rho}(\mbf{k})=\sqrt{(k-\mu)^2+\Delta_{\rho}(\mbf{k})^2}$,
and the gluon energy $\omega$ includes polarization effects
of a gluon propagating in the quark medium. The resulting high density
effective Hamiltonian is given by
\begin{eqnarray}
H_{eff} &=& \tilde{H}_0+V_{qq}
\,,\label{eq:1.8}\end{eqnarray}
where $V_{qq}$ Eq. (\ref{eq:1.6}) is also subject to the BV transformation.
$H_{eff}$ describes dynamics of quarks near the Fermi surface
in dense QCD. In order to make calculations selfconsistent we use
$\tilde{H}_{0}$ instead of $H_{0}$ in the flow equations.

Further we specify the terms in the high density effective Hamiltonian
$H_{eff}$ Eq. (\ref{eq:1.8}) or Eq. (\ref{eq:1.5}).
The kernel of the effective diquark interaction Eq. (\ref{eq:1.6})
is given by \cite{Gubankova}
\begin{eqnarray}  
V_{qq}^{\rho\rho'}(\mbf{k},\mbf{p})&=& 
-V_{dyn}^{\rho\rho'}(\mbf{k},\mbf{p})\left(\frac{3-\hat{k}\cdot\hat{p}}{2}\right)
\nonumber\\
&& -V_{inst}^{\rho\rho'}(\mbf{k},\mbf{p})\left(\frac{1+\hat{k}\cdot\hat{p}}{2}\right)
\,,\label{eq:1.9}\end{eqnarray}
where
\begin{eqnarray} 
V_{dyn}^{\rho\rho'}(\mbf{k},\mbf{p}) &=& \frac{2g^2}{3}
\frac{1}{\delta E^{\rho\rho'}(\mbf{k},\mbf{p})^2
+\omega_{M}(\mbf{k}-\mbf{p})^2}
\nonumber\\
V_{inst}(\mbf{k},\mbf{p}) &=& \frac{2g^2}{3}
\frac{1}{\omega_{E}(\mbf{k}-\mbf{p})^2}
\,.\label{eq:1.10}\end{eqnarray}
and the energy difference of in- and out-going
quarks is $\delta E^{\rho\rho'}(\mbf{k},\mbf{p})
=E^{\rho}(\mbf{k})-E^{\rho'}(\mbf{p})$.
Gluon energy, $\omega(\mbf{q})^2=\mbf{q}^2+M(\mbf{q})^2$, 
contains the Debye screening mass for electric gluon,
$M_{E}^2\sim g^2\mu^2 N_f$, and the Landau damping mass for magnetic gluon,
$M_{M}(\mbf{q})^2\sim g^2\mu^2 N_f E(\mbf{q})/|\mbf{q}|$.
In the magnetic interaction $V_{dyn}$ both factors $\delta E^2$
and Landau damping mass $M_{M}^2$ are dynamical, i.e. they depend on energies/momenta.
Dispersion relation is $|\mbf{q}|\sim E(\mbf{q})$ when $\delta E^2$ is taken into account,
and it changes to $|\mbf{q}|\sim E(\mbf{q})^{1/3}$ for Landau damping.
Effective gluon dispersion determines the numerical factor in exponent
for the gap solution.   
Dynamical magnetic interaction generated by flow equations
has the form $-1/(\mbf{q}^2+\delta E^2)$ as compared with
the equal time perturbation theory result (or interaction obtained
in the second order via Fr\"olich transformation) $-1/(\mbf{q}^2-\delta E^2)$
where the energy difference has opposite sign. The latter interaction
has an unphysical pole. To avoid it $\delta E^2$ is usually 
neglected near the Fermi surface. This is a valid argument when only 
the BCS singularity through the anomalous quark propagator is present,
and the gluon propagator $1/\mbf{q}^2$ is replaced by a point-like
interaction in the gap equation. 
However in our case of a regular interaction it is safe to keep $\delta E^2$,
which might play a role of the IR regulator 
in the collinear limit $\mbf{q}\sim 0$.   

The system of gap equations for the condensates defined in
Eq. (\ref{eq:1.6}) is given by \cite{Gubankova}
\begin{eqnarray}
\Delta_1(\mbf{p}) &=& 8G^{81}(\mbf{p})
\nonumber\\
\Delta_8(\mbf{p}) &=& G^{18}(\mbf{p})-2G^{88}(\mbf{p})
\,,\label{eq:1.11}\end{eqnarray}
where the integral in the right hand side
\begin{eqnarray}
 G^{\rho\rho'}(\mbf{p}) &=& \frac{1}{4}\int \frac{d\mbf{k}}{(2\pi)^3}
\frac{1}{2}\frac{\Delta_{\rho}(\mbf{k})}{E_{\rho}(\mbf{k})}
\label{eq:1.12}\\
&\times&V_{qq}^{\rho\rho'}(\mbf{k},\mbf{p})R^{\rho\rho'}(\mbf{k},\mbf{p};\Lambda)
+CT(\Lambda)\nonumber
\,,\end{eqnarray}
includes the anomalous $3$-d quark propagator, $\Delta_{\rho}/2E_{\rho}$,
the effective gluon propagator in the form of the effective diquark interaction,
$V_{qq}$, given in Eq. (\ref{eq:1.9}), and the UV regulating function
$R^{\rho\rho'}(\mbf{k},\mbf{p};\Lambda) =
\exp\left(-[\mbf{k}-\mbf{p}]^2/\Lambda^2\right)$.
Regulating function generated by flow equations is
$\exp\left(-[\delta E^{\rho\rho'}(\mbf{k},\mbf{p})^2
+\omega_{M}(\mbf{k}-\mbf{p})^2]/\Lambda^2\right)$,
which we approximate by $\exp\left(-[\mbf{k}-\mbf{p}]^2/\Lambda^2\right)$
for high densities. It is important that the UV regulator $R^{\rho\rho'}$ 
arises automatically from flow equations that supports an underlying concept
of flow equations being a renormalization group method. $CT(\Lambda)$ 
is the second order local counterterm chosen from the requirement
that the condesates do not depend on the UV cutoff,
$d\Delta_{i}/d\Lambda =0$ for $i=1,8$. 
Calculations by flow equations are done in the gapped
theory that is reflected by the energies $E(\mbf{k})$ containing condensates.

System of gap equations, Eqs. (\ref{eq:1.11},\ref{eq:1.12}),
can be solved only numerically.
However, an approximate analytical analyses can be done by converting 
integral gap equation into differential one \cite{Son}.
Approximate solution is given by $\Delta_{i}=\Delta_{0}^{(i)}\sin(\bar{g}x)$
with $\Delta_{0}^{(i)}=2b_{i}\mu\exp(-\pi/2\bar{g})$ for $i=1,8$,
where $x$ is a new energy variable, $\bar{g}$ is an effective coupling
proportional to $g$ and depending on effective gluon dispersion, and
$b_{i}$ are numerical factors ($b_{1}=-2b_{8}$ for color and flavor antisymmetric
condensate).      
Integral $G$ in the right hand side of the gap equation, Eq. (\ref{eq:1.12}),
has the collinear (through the gluon propagator) and the BCS 
(through the quark propagator) IR singularities.
Together these singularities give double logarithm structure for $G$
and characteristic enhanced solution $\Delta_{0}\sim\exp(-c/g)$ 
for color superconductivity instead of the BCS solution $\sim\exp(-c/g^2)$. 
BSC singularity is regulated by the condensate $\Delta_{\rho}$ 
in the energy $E_{\rho}$. 
In the case of all three quarks are massless,
the collinear singularity is cutoff by $\delta E(\mbf{k},\mbf{p})^2$ 
for the momenta close to the Fermi surface 
$|\mbf{k}-\mu|\leq\Delta$ and by the Landau damping factor $M_{M}(\mbf{k},\mbf{p})^2$ 
for the momenta away from the Fermi surface $\Delta\leq |\mbf{k}-\mu|\leq\mu$.
Landau damping of a magnetic gluon is calculated in a quark-gluon plasma,
which is a normal phase, since for large momenta gluon distingiushes
individual quarks in a Cooper pair and effectively 'sees' quark plasma rather than
a superconducting state. For massless quarks and zero temperatures the characteristic
double logaritm in $G$ which leads to enhanced solution 
is saturated by momenta away from the Fermi surface, and hence as first realized
by Son Landau damping determines the gap \cite{Son}.
Changing parameters may change numerical factors in the solution for the gap. 
In the case of massive $s$-quark, $m_u,m_d\ll m_s$, the factor $\delta E^2$
might regulate the collinear singularity contributing to the condensate $\Delta$.

Generalization to nonzero temperature is done by inserting
factor $\tanh(E_{\rho}(\mbf{k})/2T)$ into $G$, Eq. (\ref{eq:1.12}). 
We assume that the effective gluon mass
is not affected by small temperatures, since $M\sim g\mu$ and  
$T\leq \Delta\ll g\mu$ otherwise superconducting condensate will melt. 
For the momenta away from 
the Fermi surface, $\Delta\leq |\mbf{k}-\mu|\leq \mu$, $\tanh(E(\mbf{k})/2T)\sim 1$,
hence there are no temperature effects for large energies. For the momenta close
to the Fermi surface, $|\mbf{k}-\mu|\leq \Delta$, temperature distribution factor
is important. Thus only region around the Fermi surface is affected by the temperature. 
However, this region of momenta does not contribute to the double
logarithm of $G$ and hence to the enhanced gap, but rather to the BCS solution
(momenta close to the Fermi surface give single logarithm).
Therefore, temperature dependence of the condensate is of the BCS type,
i.e. $\Delta(T)\sim \Delta(T=0)\left(1-T/T_c\right)^{1/2}$ with the connection
between zero temperature condensate and the critical temperature given by 
$T_c=0.577\Delta(T=0)$.   

In conclusion, we obtained an effective high density Hamiltonian for superconductivity
which includes an effective interaction between quarks. Also we obtained the coupled
set of gap equations for the condensates of the CFL phase of massless $N_f=3$
dense QCD. The formalism used enables one to consider
a more general case of nonzero $s$-quark mass.


\begin{thebibliography}{9}


\bibitem{AlfordRajagopalWilczek} M.~Alford, K.~Rajagopal and F.~Wilczek,
{\it Nucl. Phys.}  {\bf B537}, 443 (1999).

\bibitem{Gubankova} E.~Gubankova, hep-ph/0208015.

\bibitem{Son} D.~T.~Son,
{\it Phys. Rev.}  {\bf D59}, 094019 (1999).



\end{thebibliography}
\end{document}